\documentclass{aa}  
\usepackage{graphicx}
\usepackage{txfonts}
\usepackage[colorlinks=true, allcolors=blue]{hyperref}
\usepackage{cancel}
\usepackage{placeins}
\usepackage{subcaption}   
\usepackage{lscape}

\usepackage{gensymb}
\usepackage{booktabs}
\usepackage{amsmath}
\usepackage{tablefootnote}
\usepackage{soul}
\usepackage{orcidlink}
\begin{document} 

\title{{NICER}  observations of type-I X-ray bursts from  the ultra-compact X-ray binary \mbox{M15 X-2}}
\titlerunning{X-ray bursts from ultra-compact X-ray binary \mbox{M15 X-2}}
\authorrunning{M.~A.~D\'iaz~Teodori et al.}

\author{Mar\'ia~Alejandra~D\'iaz~Teodori
\inst{1,2} \orcidlink{0009-0002-1852-7671}
\and
Jari~J.~E.~Kajava\inst{3}\orcidlink{0000-0002-3010-8333} 
\and
Celia~S\'anchez-Fern\'andez\inst{4}\orcidlink{0000-0002-0778-6048}
\and
Andrea~Sanna\inst{5}\orcidlink{0000-0002-0118-2649}
\and
Mason~Ng\inst{6,7,8}\orcidlink{0000-0002-0940-6563} 
\and
Juri~Poutanen\inst{1}\orcidlink{0000-0002-0983-0049}
}

\institute{Department of Physics and Astronomy, FI-20014 University of Turku, Finland\\
\email{madiazteo@gmail.com}
\and
Nordic Optical Telescope, 
Rambla José Ana Fernández, Pérez 7, E-38711 Breña Baja, Spain 
\and
Serco for the European Space Agency (ESA), European Space Astronomy Centre, Camino Bajo del Castillo s/n, E-28692 Villanueva de la Ca\~{n}ada, Madrid, Spain
\and
ATG Europe for the European Space Agency (ESA), European Space Astronomy Centre, Camino Bajo del Castillo s/n, E-28692 Villanueva de la Ca\~{n}ada, Madrid, Spain  
\and 
Dipartimento di Fisica, Universit\'a degli Studi di Cagliari, SP Monserrato-Sestu km 0.7, I-09042 Monserrato, Italy
\and 
MIT Kavli Institute for Astrophysics and Space Research, Massachusetts Institute of Technology, 77 Massachusetts Avenue, Cambridge, MA 02139, USA
\and 
Department of Physics, McGill University, 3600 rue University, Montr\'{e}al, QC H3A 2T8, Canada
\and 
Trottier Space Institute, McGill University, 3550 rue University, Montr\'{e}al, QC H3A 2A7, Canada}

   \date{Received XX; accepted XX}
 
  \abstract
  {Type-I X-ray bursts are thermonuclear explosions caused by the unstable burning of accreted material on the surface of neutron stars. We report the detection of seven type-I X-ray bursts from the ultracompact X-ray binary \mbox{M15 X-2} observed by the {Neutron Star Interior Composition Explorer (NICER)} during its 2022 outburst. We found all the bursts occurred in the soft state and exhibited similar light curve profiles, with no cases of photospheric radius expansion. Time-resolved spectroscopy showed clear deviations from the blackbody model during the first ten seconds of all the bursts. The fits were improved by using the enhanced persistent emission ‘$f_{\rm a}$’ method, which we interpret as evidence of burst-disk interaction. We compared the performance of these models against a neutron star atmosphere model and found it made no significant improvements. After analyzing the burst rise times and fuel composition, we propose that these bursts were powered by the burning of pure helium, confirming the ultracompact nature of the source. }

   \maketitle
%

\section{Introduction}
\label{section:intro}
Since their discovery in 1976 \citep{Grindlay_1976}, type-I X-ray bursts (hereafter, X-ray bursts) have been a key tool in the study of neutron stars  (NSs),  offering insights into both their accretion processes and intrinsic characteristics. These bursts are thermonuclear explosions that take place in low-mass X-ray binary (LMXB) systems with a NS primary.  They are caused by the accretion of matter from the low-mass companion onto the NS, which creates a fuel layer on the NS surface that grows and compresses until the conditions for thermonuclear ignition are reached 
\citep[see reviews  in][]{lewin_993,strohmayer_2006,galloway_2021}.

X-ray bursts are manifested as a sudden increase in X-ray emission followed by an exponential
decay, usually lasting $\sim$10--100~s \citep{lewin_993}. In rare instances, there are bursts of longer duration: intermediate-duration bursts last from minutes to an hour, and superbursts can last many hours \citep[e.g.,][]{strohmayer_2002,cumming_2006}. The burst duration is related to the depth and composition of the fuel layer. Moreover, both the H/He ratio \citep{fujimoto_1981,galloway_2021} and the metallicity \citep{Heger_2007} impact the burst timescale, since both of these properties affect the nuclear-burning pathways.

Through time-resolved spectroscopy of X-ray bursts,  insights can be gained into the properties of the NS and its accretion environment. 
The spectrum of the system during an X-ray burst is usually modeled by the sum of the persistent emission spectrum, which is assumed to not vary during the burst \citep[e.g.,][]{lewin_993}, and a blackbody component, with evolving parameters to describe the burst itself. 

During the initial fuel ignition, the measured blackbody temperature increases to a maximum of 2--3 keV and then gradually decreases as the flux drops and the photosphere cools down. Nonetheless, numerous bursts show deviations from the traditional model \citep{vanParadijs_1990}, either due to changes in the persistent emission \citep{worpel_2013,worpel_2015} or the intrinsic burst spectrum \citep[e.g.,][]{Kajava_2017a,sanchezfernandez_2020}. The radiation produced during X-ray bursts may interact with the accretion disk. In this context, several phenomena that could affect the persistent emission have been studied, namely: radiatively driven outflows \citep[e.g.,][]{Denegaar_2013}, increased inflow rate due to the Poynting-Robertson drag \citep{worpel_2013}, and changes in the disk structure due to burst heating \citep[e.g.,][]{Ballantyne_2005,intZand_2011}. To account for enhanced persistent emission, \citet{worpel_2013} and \citet{intZand_2013} proposed the  \textit{``$f_{\rm a}$''}  method, which involves adding a multiplicative factor \textit{$f_{\rm a}$} to the persistent emission model when fitting the spectra. \citet{worpel_2013, worpel_2015} applied this method to a large sample of bursts and found enhanced persistent emission to be common. This method, however, assumes that the shape of the persistent emission does not change during the burst. Regarding the intrinsic burst spectrum, the emission from hot NS atmospheres is not expected to be an ideal blackbody: the combined effect of Compton scattering and free-free (and bound-free) emission and absorption in the NS atmosphere modifies the spectrum into a diluted blackbody  \citep[e.g.,][]{London_1986,suleimanov_2012}. 

\mbox{M15 X-2} is one of the three known X-ray sources in the globular cluster (GC) M15, along with AC 211 \citep{white_2001} and M15 X-3 \citep{Heinke_2009}. \mbox{M15 X-2} and AC 211 are persistent sources \citep{Heinke_2024}, whereas M15 X-3 is a faint transient.
\mbox{M15 X-2} is also an ultracompact X-ray binary (UCXB) system. UCXBs are defined by their short orbital periods ($\lesssim$60--80 min) that cannot fit hydrogen-rich companions \citep{Nelson_1986} and instead have donors ranging from white dwarfs (WDs) to helium stars to evolved main-sequence stars \citep[e.g.,][]{Yungelson_2002,Nelemans_2006}.

\begin{figure}
\centering
\includegraphics[width=\hsize]{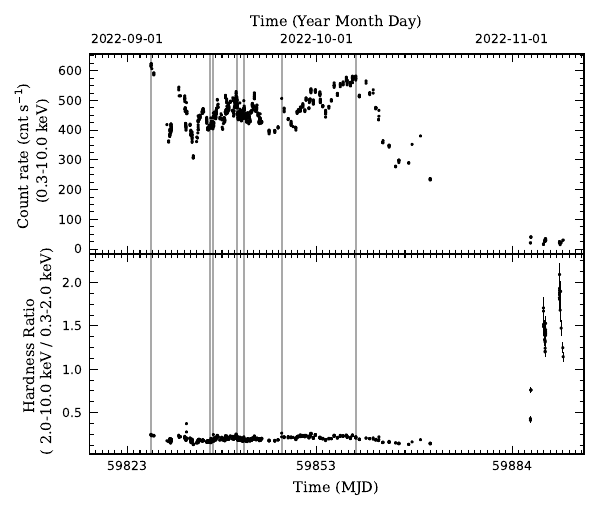}
\caption{Properties of \mbox{M15 X-2} outburst detected by NICER in the 0.3--10 keV energy band. 
\textit{Top panel}: the light curve. 
\textit{Bottom panel}: Hardness ratio of the count rates between the 2--10 and 0.3--2.0~keV energy bands. 
The X-ray bursts have been removed from the light curve, but the detection time of each burst is marked with a grey vertical line. For the fifth burst, only the tail was observed, so the onset is an approximation based on Fig.~\ref{fig:Burst_example_LC}.}

\label{fig:OutburstLC}
\end{figure}

The discovery of \mbox{M15 X-2} is closely tied to AC~211 and, in fact, it helped resolve the decade-long mystery surrounding AC~211. The LMXB AC~211  is known for its edge-on inclination, obscuring the central X-ray source behind the accretion disk. This became a source of confusion in 1988, when the \textit{Ginga} satellite detected a very potent X-ray burst \citep{vanParadijs_1990} from AC 211, an event that should have been concealed by the disk. Similarly,  the Rossi X-ray Timing Explorer (\textit{RXTE}) detected another powerful burst in 2000 \citep{Smale_2001}. 
Eventually, \citet{white_2001},  through \textit{Chandra} observations, discovered that the bursts were actually coming from \mbox{M15 X-2}, which was 2\farcs7 away from AC 211 and had been previously unresolved. 
 
Four years later, \citet{dieball_2005} determined the period of the far-ultraviolet (FUV) counterpart of \mbox{M15 X-2} to be $\sim$23~min, confirming it as a UCXB. Additionally, they found FUV evidence of C and/or He lines. Later on, \citet{koliopanos_2020} studied the X-ray spectrum of \mbox{M15 X-2} and found no Fe K$\alpha$ line emission, which they concluded was evidence of a C/O or O/Ne/Mg WD donor. Thus, the hydrogen-poor nature of the companion was confirmed, but its exact classification remained unclear. 

Since its identification in 2001, bursts from \mbox{M15 X-2} have been observed again in 2003 by \textit{BeppoSAX}/WFC \citep{kuulkers_2003}, as well as in 2011 by the Monitor of All-sky X-ray Image  \citep[{MAXI};][]{atel_millerjones_evla_2011,atel_sivakoff_chandra_2011} and in 2013 again by {MAXI} \citep{atel_tomida_maxi_2013,atel_pooley_chandra_2013}. Additionally, \citet{charles_2002} identified 15 burst candidates in archival All Sky Monitor/\textit{RXTE} data. 

On 2022 August 26, after almost a decade in a low state, a new outburst from the M15 field was detected by {MAXI}/Gas Slit Camera  \citep{atel_ng_nicer_2022}. On 2022 September~4,  the Neutron Star Interior Composition Explorer ({NICER}) performed a 1.6~ks observation of M15 and observed one X-ray burst.  
On 2022 September~27, \textit{Chandra}/ACIS-S conducted follow-up observations and concluded \mbox{M15 X-2} was the source of the outburst and subsequent X-ray burst \citep{atel_homan_chandra_2022}. {NICER} continued carrying out follow-up observations until 2022 November 9.

In this work, we report the analysis of seven X-ray bursts from \mbox{M15 X-2} observed by {NICER} during its 2022 outburst. In Sect.~\ref{section:observations}, we describe the observational data and reduction procedure. In Sect.~\ref{section:analysis}, we present the spectral and timing analysis results. In Sect.~\ref{section:discussion}, we discuss the results obtained and in Sect.~\ref{section:conclusions}, we summarize the main conclusions of the study. 

\section{Observations}
\label{section:observations}

\begin{table}
\caption{Basic information on the seven bursts.}
\centering
\begin{tabular}{clc} 
\toprule
\toprule
Burst & ID & Burst Onset (MJD)\\ \midrule
1& 5034140101&59826.73942   \\ 
2& 5034140110&59836.09463\\ 
3& 5034140110&59836.61754 \\  
4& 5034140114&59840.43209 \\ 
5& 5034140115\tablefootmark{a} &59841.45299 \\ 
6& 5034140121&59847.46255\\ 
7& 5034140131&59859.19994\\ \bottomrule
\end{tabular}
\tablefoot{Time is in TT system.  \tablefoottext{a}{The onset is approximate, based on Fig.~\ref{fig:Burst_example_LC}, because only the tail of this burst was observed.}}
\label{tab:basics}
\end{table}

We analyzed the 51 \mbox{M15 X-2} {NICER} observations (ObsIDs 5034140101-5034140152) that monitored the outburst between August and November 2022 (see Fig.~\ref{fig:OutburstLC}). The total coverage was 146.8~ks. We found seven bursts, with the corresponding ObsIDs and burst onset times given in Table~\ref{tab:basics}. 

We processed the data using $\textsc{heasoft}$ v 6.31.1 and the {NICER} Data Analysis Software $\textsc{nicerdas}$.  This was done through the use of $\textsc{HEASoftPy}$, a \textsc{Python 3} package, which gives access to the $\textsc{heasoft}$ tools.
The processing consists of several steps, with the first being the use of standard {NICER}-recommended calibration and filtering tool \textsc{nicerl2} to produce the clean event files. Standard screening criteria were applied, which include a pointing offset of $<54\arcsec$, Earth limb elevation angle of $>15\degr$,  elevation angle with respect to the bright Earth limb of  $>30\degr$,  being outside of the South Atlantic Anomaly, and an overshoot rate of $<1 \mathrm{cnt}\,\mathrm{s}^{-1}$ per detector.  Since the \textsc{heasoft} v6.31. release, \textsc{nicerl2} requires updated geomagnetic quantities; in this case, we have used data from NOAA's {Space Weather Prediction Center}. 

From the event files, the light curves were produced using the \textsc{heasoft} task  \texttt{extractor} with 1\,s binning. It was checked beforehand that all 52 NICER detectors were enabled throughout the observations. Bursts were detected by performing sigma-clipping with a 5$\sigma$ limit. In the intervals where a burst was detected, the onset time was obtained through the following procedure: 1) finding the peak of the burst; 2) calculating the average persistent emission by averaging between $-40$\,s and $-20$\,s before burst peak time (this timescale ensured we could study bursts, e.g., Burst 3, which occurred $< 50$\,s after the observation onset); 3) subtracting the persistent emission to the light-curve; 4) dividing the light curve by the peak value of the burst, such that it was now normalized; and 5) starting from the peak, searching backward until we found the first point where the amplitude went below 0.1, that being the onset time. If there is not enough data to go back 40\,s, we averaged the first 5\,s of the observation. 

To determine the rise and decay timescales of the burst, the individual light curves were fitted by applying the phenomenological model proposed by \citet{Pike_2021}. This work uses a piecewise approach in which a fast-rise exponential-decay (FRED) curve models the rise, peak, and beginning of the decay, and the later decay is modeled as a power law. This allows us to model the X-ray burst signature quick rise and slower decay, while also taking into consideration the later decay of X-ray bursts  more closely resembles a power law, as shown, for example, by \citet{intZand_2017}. The model is expressed as:
\begin{equation} \label{eq:lc_model}
f(t) = 
\begin{cases} 
\displaystyle 
A \exp\left[ -\frac{\tau_{\rm R}}{t} - \frac{t}{\tau_{\rm D}} \right] + C, & t \leq t_{\rm tail}, \\ 
\displaystyle  
B \left( \frac{t }{t_{\rm tail} } \right)^{-\gamma} + C, & t > t_{\rm tail}, 
\end{cases}
\end{equation}
where $A$ is the peak intensity, $\tau_{\rm R}$ is the rise timescale, $\tau_{\rm D}$ is the decay timescale, $t_{\rm tail}$ is the transition time from exponential to the power-law tail, $\gamma$ is the power-law index, and $B \equiv A \exp\left[ -\frac{\tau_{\rm R}}{t_{\text{\rm tail}} - t_0} - \frac{t_{\rm tail} - t_0}{\tau_{\rm D}} \right]$, which ensures the exponential and the power-law tails have the same count rate at $t=t_{\rm tail}$.

The burst spectra were also produced with \texttt{extractor}. For the pre-burst spectra, we extracted 20~s intervals ending at least 20~s before burst onset, the only exception being burst \#5, as the onset of the burst happened before our observation started, so the 20~s interval was extracted from the previous observation, approximately $10^4$~s before the start of the burst. To perform time-resolved spectroscopy of the burst itself, dynamic intervals containing 1500 counts were generated, and spectra were extracted for each interval. The spectrum files were then binned using the \texttt{grppha} tool, with a minimum condition of 20 counts per bin to ensure we are in the domain of chi-square statistics.

To produce the response matrix files (RMF) and ancillary response files (ARF) for the spectral analysis, the \textsc{nicerdas} tools \texttt{nicerrmf} and \texttt{nicerarf} were used, respectively. The background was estimated using the \texttt{nibackgen3C50} task \citep{Remillard_2022}. For this analysis, the NICER calibration files were obtained through the \texttt{CALDB} remote access,  and the October 2022 version was used.  Once all the necessary files were created, the spectra were analyzed using the \textsc{python} interface of \textsc{xspec} \citep{Arnaud_1996}, \textsc{PyXspec}.For this purpose, \textsc{xspec} v12.13.0 was used. The models used to analyze the spectra will be explained in detail in Sect.~\ref{section:analysis}. Uncertainties were calculated at 1$\sigma$ confidence level.
 
\section{Analysis and results} 
\label{section:analysis}

\subsection{Outburst analysis}

The outburst light curve and hardness plot are presented in Fig.~\ref{fig:OutburstLC}, encompassing all the observations taken between August and November 2022. The X-ray bursts have been removed from the light curve, but the onset times are marked with vertical grey lines. Six X-ray bursts, plus the tail of the seventh, were identified during this outburst. 

In the bottom panel of Fig.~\ref{fig:OutburstLC}, we see how the hardness ratio increases significantly near MJD~59889. We extracted spectra for these pointings and noticed that they can be modeled using a \textsc{diskbb} model of $T_{\rm in} \approx 2$~keV, with partial covering absorber of $N_{\rm H} \sim 10^{22}\,{\rm cm^{-2}}$, a partial covering fraction of about $PCF \approx 0.92$, and 0.3--10~keV flux of $F_{\rm x} \sim 10^{-10}\,{\rm erg\,cm^{-2}\,s^{-1}}$.
These values are very similar to earlier measurements of the nearby AC~211 \citep{Sidoli_2000,white_2001}, suggesting that at the end of the NICER observing campaign, AC~211 is already brighter than \mbox{M15 X-2}. If we consider that the outburst lasted until the last observation in the softer state, the duration of the outburst was 54~d.

\begin{table*}
\caption{Best-fit parameters of the model \texttt{tbabs*(diskbb+comptt)} for  pre-burst spectra of the seven bursts. }
\centering
\begin{tabular}{ccccccc} 
\hline\hline
  Burst  & \texttt{tbabs} $N_{\rm H}$  & 
  \texttt{diskbb} $T_{\rm in}$ & 
 \texttt{diskbb} norm   &  \texttt{compTT} norm &  Flux & $\chi^2$/dof \\
& ($10^{20}$~cm$^{-2}$) &  (keV) & && ($10^{-10}$~erg~cm$^{-2}$~s$^{-1}$)\\   
  \hline
         1 & $5.0 \pm 0.4$ & $0.80 \pm 0.03$ & $110 \pm 10$ & $0.043 \pm 0.003$ & $14.4\pm 0.1$ & 0.86\\ 
         2 & $6.4 \pm 0.5$ & $0.60 \pm 0.02$ & $200 \pm 30$ & $0.021 \pm 0.002$ & $8.0 \pm 0.1$ & 0.98\\ 
         3 & $4.9 \pm 0.5$ & $0.66 \pm 0.02$ & $142 \pm 20$ & $0.025 \pm 0.002$ & $8.5 \pm 0.1$ & 1.07\\  
         4 & $5.5 \pm 0.6$ & $0.74 \pm 0.02$ & $110 \pm 10$ & $0.030 \pm 0.002$ & $10.5\pm 0.1$ & 0.87\\ 
         5 & $5.7 \pm 0.5$ & $0.70 \pm 0.02$ & $130 \pm 20$ & $0.023 \pm 0.002$ & $9.1 \pm 0.1$ & 0.98\\ 
         6 & $6.1 \pm 0.5$ & $0.72 \pm 0.02$ & $120 \pm 10$ & $0.026 \pm 0.002$ & $9.8 \pm 0.1$ & 1.06\\ 
         7 & $6.0 \pm 0.4$ & $0.74 \pm 0.02$ & $140 \pm 20$ & $0.035 \pm 0.003$ & $12.7\pm 0.1$ & 0.95\\
\hline
\end{tabular}
\label{tab:my_label}
\end{table*}

We studied the persistent emission before each burst by extracting 20\,s intervals ending at least 20\,s before burst onset. All of the spectra were fit to the model \texttt{tbabs*(diskbb+comptt)}. This consists of a multicolor
disk blackbody component \citep[\texttt{diskbb},][]{Makishima_1986}, 
combined with a Comptonization component  \citep[\texttt{comptt},][]{Titarchuk_1994} and modified by the Tübingen–Boulder interstellar absorption model \texttt{tbabs}. We used the element abundances provided by \citet{Wilms_2000}. 

For the \texttt{diskbb} component, we allowed both of its parameters, the inner disk radius temperature, $T_{\rm in}$, and the normalization, $K_{\rm diskbb}$, to vary.
We fixed the \texttt{comptt} seed temperature $T_0 = 1.5$~keV, electron temperature $T_{\rm e} = 3.3$~keV and the optical depth  $\tau = 5.0,$ following \citet{RG2006}. Only the normalization $K_{\rm comptt}$ was allowed to vary.
The fitting results for each burst are displayed in Table~\ref{tab:my_label}. The average values obtained for the parameters are $N_{\rm H} = (5.6 \pm 0.6) \times 10^{20}$~cm$^{-2}$, $T_{\rm in} = 0.69  \pm 0.05$~keV, $K_{\rm diskbb} = 140 \pm 30$, and $K_{\rm comptt} = 0.029 \pm 0.007$.

\begin{figure}
\centering
\includegraphics[width=\hsize]{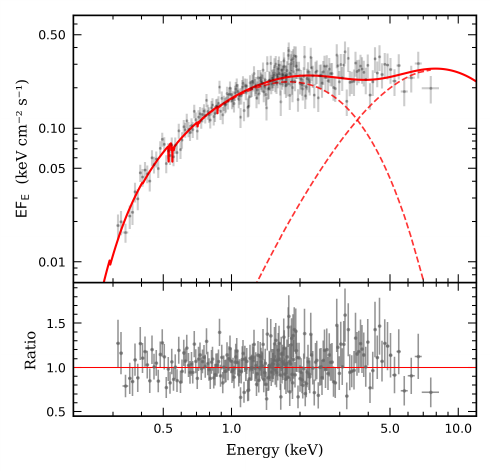}
\caption{X-ray spectrum of the pre-burst emission for burst \#4. 
Spectra of other bursts are shown in Fig.~\ref{fig:PreBurst_spectrum1}.
The solid red line represents the total model \texttt{tbabs*(diskbb+comptt)}, while the dashed lines represent  \texttt{diskbb} and \texttt{comptt} components separately, and the gray points show the data. 
The pre-burst emission was chosen as a 20~s interval taking place 20~s before the burst. }
\label{fig:PreBurst_spectrum}
\end{figure}

From the average disk normalization, we estimated the apparent inner disk radius $R_{\rm in}$, which is related to each other as $K_{\rm diskbb}=(R_{\rm in}/D_{10})^2 \cos \theta$, where $D_{10}$ is the distance to the source in units of 10 kpc, and $\theta$ the disk inclination \citep{Makishima_1986}. Using the Gaia EDR3 value of 10.7 kpc for the distance \citep{Baumgard_2021}  and taking $\theta= 34\degr$   \citep{dieball_2005}, an average $R_{\rm  in}$ $\approx$ 11~km is obtained. This is close to a NS radius, which is thought to be between 10 to 15~km.

The persistent spectrum of burst \#4 is shown in Fig.~\ref{fig:PreBurst_spectrum}.
The model seems to fit the data with no significant deviations at any point in the energy range. The obtained parameter values are compatible with the soft state \citep{done_2007}.

\begin{table*}
\caption{Best-fit rise timescale, decay timescale and power-law value using Eq.~\eqref{eq:lc_model} to the burst profiles.}
\centering
\begin{tabular}{cccc} 
\hline  \hline 
Burst & $\tau_{\rm R}$ & $\tau_{\rm D}$ & $\gamma$  \\ 
& (s) & (s)   \\
 \hline 
        1& 0.0791 $\pm$ 0.0004  & 15  $\pm$ 1 & 1.0770 $\pm$ 0.0008 \\ 
         2& 0.1289 $\pm$ 0.0001  & 30  $\pm$ 5 & 1.0262 $\pm$ 0.0001 \\
         3& 0.1434 $\pm$ 0.0001   & 28  $\pm$ 6 & 0.9752  $\pm$ 0.0001 \\ 
         4&  0.04201 $\pm$ 0.00004  & 15.0  $\pm$ 0.8 & 0.9687 $\pm$ 0.0001 \\
         6& 0.08954 $\pm$ 0.00006 & 22  $\pm$ 5 & 0.9754 $\pm$ 0.0001 \\ \hline
    \end{tabular}
\tablefoot{Burst 5 is omitted because only the tail was observed.}
\label{tab:lcs}
\end{table*}

\subsection{Burst light curves}

\begin{figure}
\centering
\includegraphics[width=\hsize]{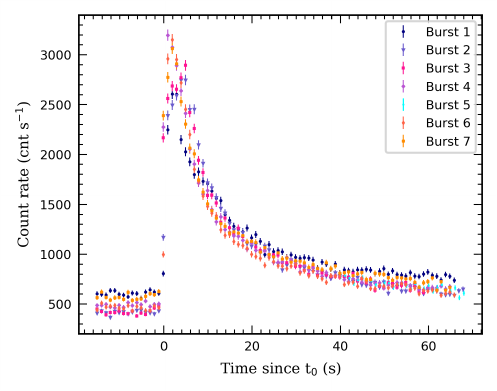}
\caption{Overlapped light curves of the seven \mbox{M15 X-2} bursts in the 0.3--10 keV band  with 1\,s binning. The light curves are aligned by the burst onset ($t_0$). Each color and symbol represents a different burst. For burst 5, only the tail was observed, so it was manually positioned to match the morphology of the other bursts.
}
\label{fig:Burst_example_LC}
\end{figure}
\begin{figure}
\centering \includegraphics[width=\hsize]{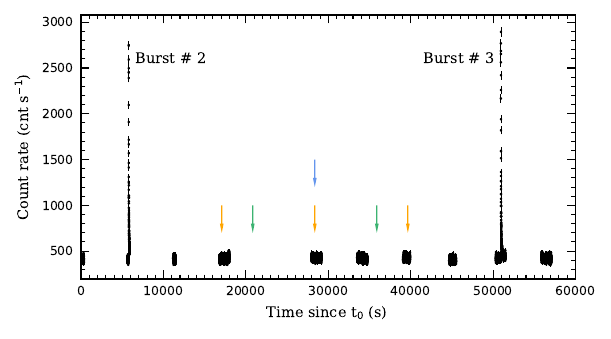}
\caption{Cropped light curve of ObsID 5034140110, showing two consecutive X-ray bursts (\#2 and \#3). Assuming the recurrence time is the interval between the two bursts, the blue arrow indicates where a burst would appear with half the recurrence time, the green arrows where they would appear with a third of the recurrence time, and the orange arrows where they would appear with a fourth of the recurrence time. Time is relative to the onset of the observation $t_0$= 59859.1880 MJD (TT).}%
\label{fig:doubleLC}
\end{figure}

Individual light curves of the seven X-ray bursts were extracted in the 0.3--10~keV energy band with 1~s binning. In Fig.~\ref{fig:Burst_example_LC}, the light curves are displayed overlapping each other, aligned by the burst onset. We see that the bursts display similar morphology, with the count rate peaking at around 2500--3000 cnt\,s$^{-1}$. 

The results of the fittings of model Eq.~\eqref{eq:lc_model} to the burst light curves are presented in Table~\ref{tab:lcs}.
From the derived timescales, we can get the average value of the rise, $\tau_{\rm rise} =0.10\pm0.04$~s, and of the decay time, $\tau_{\rm dec} =   22 \pm 7$~s.
We also see that all bursts have occurred around the same hardness level (see bottom panel of Fig.~\ref{fig:OutburstLC}). 

It is worth noting that the second and third bursts were observed in a single observation (ObsID 5034140110). In Fig.~\ref{fig:doubleLC}, we show a cropped version of this observation to show the two consecutive bursts. The time between the two bursts is $45179 \pm1$~s ($\sim 12.5$~h). Additionally, looking at Fig.~\ref{fig:OutburstLC} again, we see that the fourth and fifth bursts are observed quite close to each other, with the time between them being $88207\pm1$~s ($\sim 24.5$~h). 

Our findings suggest that the recurrence time is unlikely to be half or a quarter of this period, as indicated by the blue and orange arrows in Fig.~\ref{fig:doubleLC}, respectively. These arrows mark telescope pointings during these times, and no bursts were detected. The only scenario that cannot be entirely dismissed is a recurrence time of one-third of the observed interval, shown by the green arrows, due to a lack of pointings during these intervals. However, the fourth and fifth bursts, observed in close succession (Fig. \ref{fig:OutburstLC}), occurred $\sim$24.5~h apart, almost double the proposed recurrence time.

\subsection{Time-resolved burst spectroscopy}

Time-resolved spectroscopy was performed for each burst using several models. 
The results for each individual model are discussed below. In all cases the unabsorbed bolometric flux was calculated using the \texttt{cflux} model component in \textsc{xspec}.

\subsubsection{Blackbody model} 

We first attempted to model the spectra of all the bursts by adding a blackbody component to the pre-burst spectrum model, holding constant all the parameters except those of the blackbody  (i.e., assuming that the pre-burst spectrum does not change during the burst).
Thus, the used model was \texttt{tbabs*(bbody+diskbb+comptt)}.  
The values of $\chi^2$ during the first ten seconds tell us that this model does not properly fit the data. This behavior repeats in all the bursts.    

\begin{figure}
\centering
\includegraphics[width=0.9\hsize]{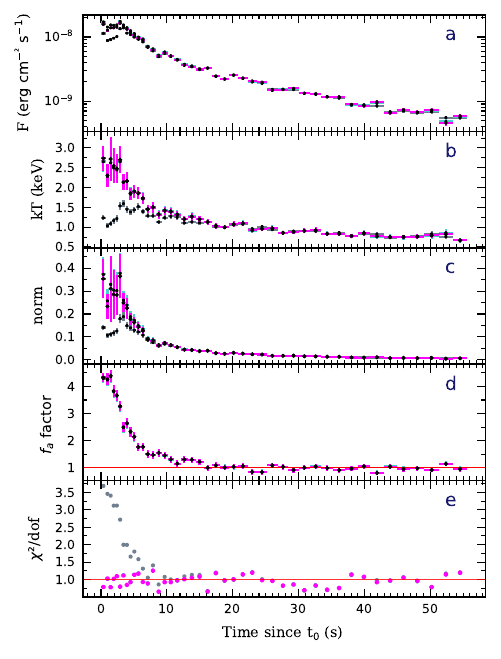}
\caption{Time-resolved spectroscopy using the $f_{\rm a}$ model for burst \#4. 
Results for other bursts are shown in Fig.~\ref{fig:blackbody_fa1}.
Gray markers are used for the blackbody model, and magenta markers are used for the first $f_{\rm a}$ case, \texttt{constant*(diskbb+comptt)}, and cyan markers for the second $f_{\rm a}$ case, \texttt{constant*diskbb+comptt}. Since their behavior is similar, the magenta markers overlap the cyan ones.
Panel (a) shows the estimated bolometric flux, panel (b) shows the time evolution of the blackbody temperature, panel (c) shows the time evolution of the blackbody normalization, panel (d) shows the evolution of the $f_{\rm a}$ factor, and panel (e) gives the reduced $\chi ^2$ fit statistic. Time is relative to the burst onset ($t_0$).}
\label{fig:blackbody_fa}
\end{figure}


\begin{figure}
\centering
\includegraphics[width=0.9\hsize]{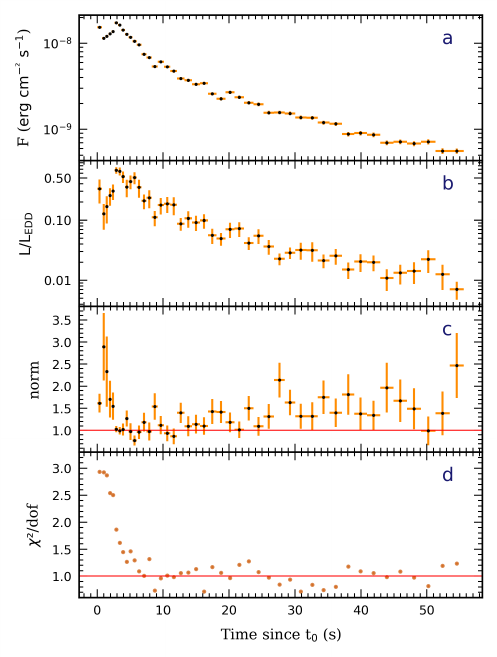}
\caption{Time-resolved spectroscopy using the NS atmosphere model for burst \#4. 
Results for other bursts are shown in Fig.~\ref{fig:burstatmo1}.
Panel (a) shows the estimated bolometric flux, panel (b) shows the time evolution of the ratio of luminosity to the  Eddington luminosity, panel (c) shows the time evolution of the normalization, and panel (d) gives the reduced $\chi^2$ fit statistic. Time is relative to the burst onset ($t_0$).}
\label{fig:burstatmo}
\end{figure}
 
To improve the fitting, we tried incorporating a free scaling ($f_{\rm a}$) factor to the persistent emission components to account for a possible increase of persistent emission during the burst, as proposed by \citet{worpel_2013,worpel_2015}. The $f_{\rm a}$ factor is implemented as a multiplicative constant in \textsc{xspec} that is free to vary during the fitting. Two cases were studied, \texttt{constant*(diskbb+comptt)}, and \texttt{constant*diskbb+comptt}. 
 
We compare in Fig.~\ref{fig:blackbody_fa} the results obtained using the $f_{\rm a}$ models with those obtained when the persistent emission is considered constant during the bursts. 
The plot is shown for burst \#5, but the same behavior is observed in all the bursts. The figure shows that both $f_{\rm a}$ models improve the fit with respect to the persistent emission model during the first 10 seconds of the burst. We note, however, that our models do not allow the selection of one $f_{\rm a}$ model over the other.   

\subsubsection{Neutron star atmosphere model}
We also attempted to use a more detailed description of the NS atmosphere and fit the burst spectra using the NS atmosphere model \texttt{burstatmo} proposed by \citet{suleimanov_2011, suleimanov_2012}.
This model takes into account  Compton scattering in the NS atmosphere, which modifies the burst spectrum to something close to a diluted blackbody. 
The model used was then \texttt{tbabs*(burstatmo+diskbb+comptt)}. 
Parameters of  \texttt{diskbb+comptt} were fixed at the pre-burst values.
The parameters of the \texttt{burstatmo} model are the radius ($R$), the ratio of the luminosity to the Eddington luminosity ($L/L_{\rm Edd}$), the distance in kpc, the logarithm of surface gravity ($\log g$), the chemical composition, and the normalization. 
The model is computed on a grid of 18 combinations of chemical composition and gravities:  six chemical compositions (pure hydrogen $X=1$, pure helium $Y=1$, and four sets of a solar mix of hydrogen and helium with various heavy element abundances) and three gravities $\log g$: 14.0, 14.3, and 14.6. 
For each combination, 26--28 models are computed for different Eddington ratios (i.e., luminosities in units of the Eddington one). 
While this grid constrains the flexibility of the model, it is sufficient to cover the typical range of neutron star masses and radii. 
In \citet{nattila_2017}, this model was used to perform NS mass-radius measurements from the evolution of X-ray burst spectra and the radius of the NS in 4U 1702$-$429 was estimated to be $R = 12.4 \pm 0.4$~km ($68\%$ credibility).

\begin{figure}
\centering
\includegraphics[width=0.9\hsize]{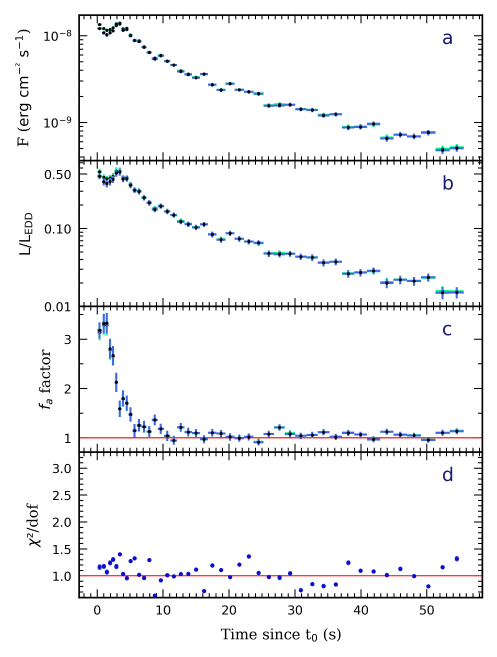}
\caption{Time-resolved spectroscopy using the NS atmosphere model in addition to the $f_{\rm a}$ correction for burst \#4. 
Results for other bursts are shown in Fig.~\ref{fig:burstatmo_fa1}.
Blue markers are used for the first $f_{\rm a}$ case, \texttt{constant*(diskbb+comptt)} and light green markers for the second $f_{\rm a}$ case, \texttt{constant*diskbb+comptt}. Since their behavior is similar, the blue markers overlap the green ones. Panel (a) shows the estimated bolometric flux, panel (b) shows the time evolution of the ratio of luminosity to  Eddington luminosity, panel (c) shows the evolution of the $f_{\rm a}$ factor, and panel (d) gives the reduced $\chi^2$ fit statistic.}
\label{fig:burstatmo_fa}
\end{figure}

For this study, the Eddington ratio and the normalization were the only parameters used for the fitting. 
Regarding the other parameters, a distance of 10.7~kpc is assumed \citep{Baumgard_2021} and the chemical composition was set to pure  helium. 
After testing different combinations of $R$ and $\log g$, we obtained the best-fit results using $R$=14.0~km and $\log g$=14.6, which are then used for the fitting. 
The fitting results are represented in Fig.~\ref{fig:burstatmo}.

The figure shows that the model still does not yield good fits for the first 6\,s of the burst if the persistent emission is tied to pre-burst values. 
We also note that the normalization deviates significantly from unity, especially during the first 6\,s, which would indicate strong anisotropy of the burst emission and is likely unphysical. 
This behavior appears in all the other bursts (see  Fig.~\ref{fig:burstatmo1}).

\begin{figure}
\centering
\includegraphics[width=\hsize]{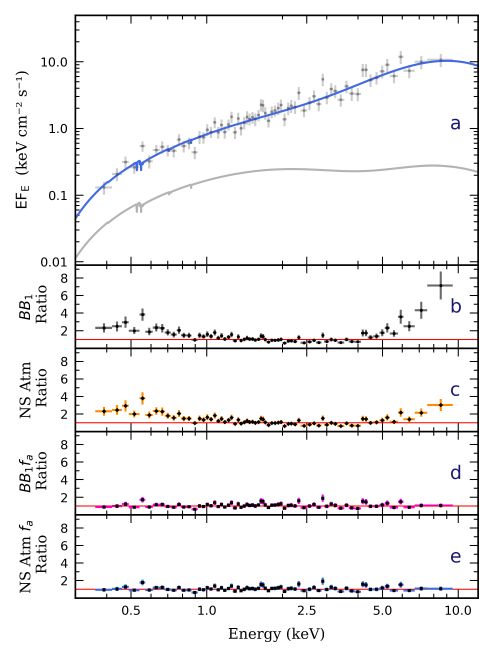}
\caption{Burst spectral energy distribution and residuals of the first second of burst \#4 when fitted by the four used models. 
Results for other bursts are shown in Fig.~\ref{fig:Residuals1}.
Panel (a) shows the spectrum, with the continuous blue line representing the NS atmosphere with $f_{\rm a}$ factor model, the continuous gray line representing the pre-burst emission model shown in Fig.~\ref{fig:PreBurst_spectrum}, and the gray points representing the data. 
Panel (b) shows the residuals for the blackbody model.
Panel (c) shows the residuals for the NS atmosphere model.
Panel (d) shows the residuals for the blackbody with $f_{\rm a}$ factor model, with magenta for the \texttt{constant*(diskbb+comptt)} case and cyan for \texttt{constant*diskbb+comptt} case, both with square markers. Since their behavior is similar, the magenta markers overlap the cyan ones.
Panel (e) shows the residuals of the NS atmosphere with $f_{\rm a}$ factor model.
}
\label{fig:Residuals}
\end{figure}
   
To improve the fitting, we tried incorporating a free scaling ($f_{\rm a}$) factor to the pre-burst spectrum when fitting the burst spectra using the NS atmosphere model. Two cases were studied, \texttt{constant*(diskbb+comptt)} and \texttt{constant*diskbb+comptt}. To do this, we fixed the \texttt{burstatmo} normalization to one. 
The results for burst \#4 are shown in Fig.~\ref{fig:burstatmo_fa}. 
This figure shows both $f_{\rm a}$ models correct the fit. 
This behavior appears in all the other bursts (see Fig.~\href{https://zenodo.org/records/14618966}{A.4}).

Figure~\ref{fig:Residuals} compares the residuals of the first second of burst \#4 for the four different models (see Fig.~\ref{fig:Residuals1} for other bursts).
This figure corroborates how the $f_{\rm a}$ models improve the fitting in both scenarios, and also shows the NS atmosphere models give slightly better residuals than the blackbody, but not enough to get a good fit.

\section{Discussion}
\label{section:discussion}

This study analyzes seven \mbox{M15 X-2} bursts detected by NICER during its 2022 outburst. We observed a significant spectral deviation from the blackbody model in the first few seconds of all the bursts and tested various models to improve the spectral fits. We also constrained the burst recurrence time, which will allow us to estimate the fuel composition. We now discuss these findings.

\subsection{Spectral state of the outburst}

All the \mbox{M15 X-2} bursts seem to have occurred in the soft spectral state, as can be inferred from the spectra of their persistent emission (Fig.~\ref{fig:PreBurst_spectrum}) and outburst hardness evolution (Fig.~\ref{fig:OutburstLC}).
This contrasts with previous bursts from this source; the 1989 and 2011 outbursts seem consistent with the hard state \citep{vanParadijs_1990,atel_millerjones_evla_2011}. 
The features of these bursts are also notably different from previous ones, with the \citet{vanParadijs_1990} and \citet{Smale_2001} bursts being particularly powerful and displaying photospheric radius expansion (PRE). 
Thus, it is plausible that this new set of bursts occurred in a different accretion regime.

\subsection{Burst spectral evolution}

Time-resolved spectroscopy indicates that all the bursts analysed in this work exhibit enhanced persistent emission during the initial seconds since burst onset. In recent years, NICER's sensitivity to soft X-rays has enabled the detection of many such cases \citep[e.g.,][]{keek_2018a,keek_2018b,Jaisawal_2019,bult_2029, guver_2022b, guver_2022a,bult_2022,yu_2024, lu_2024}. However, this is the first time bursts from this source have been observed with such low-energy coverage. 

Figures~\ref{fig:blackbody_fa} and \ref{fig:blackbody_fa1} shows that the $f_{\rm a}$ factor improves the blackbody model fitting of the first ten seconds. The peak values of $f_{\rm a}$ that we derive in this work are consistent with those found by \citet{worpel_2015} for non-PRE bursts. After those initial ten seconds, $f_{\rm a}$ returns to unity, which is consistent with previous studies although the precise duration seems to vary in each case  \citep[e.g.,][]{bult_2022}. 

This enhanced persistent emission, indicated by the $f_{\rm a}$ correction, is typically interpreted as resulting from an increased accretion rate due to Poynting-Robertson drag \citep{worpel_2013, worpel_2015}. An alternative interpretation proposed by \cite{Koljonen_2016} and \citet{Kajava_2017a}  for bursts in the soft state is that there may be a broadening of the spectrum of the spreading layer between the NS surface and the accretion disk. This broadening could resemble an increased accretion rate.

\subsection{NS atmosphere models} 

By fitting the spectra using the NS atmosphere models, we examined whether the blackbody deviations could be explained solely by the bursts' intrinsic spectra. 
Successful fits with NS atmosphere models can provide accurate mass-radius measurements, as demonstrated by \citet{nattila_2017}. 
However, Figs.~\ref{fig:burstatmo} and \ref{fig:burstatmo1}{A.3} show that chi-square values during the first ten seconds of the burst remain poor. 

Similar to the blackbody model, the fit improves when the $f_{\rm a}$ correction is applied, as seen in  Figs.~\ref{fig:burstatmo_fa} and \ref{fig:burstatmo_fa1}. The residuals in Figs.~\ref{fig:Residuals} and \ref{fig:Residuals1} indicate no notable difference between the blackbody and NS atmosphere $f_{\rm a}$ corrections. We note that in this case, $f_{\rm a}$ returns to unity in around seven seconds, slightly quicker than with the blackbody model, suggesting that the NS atmosphere model might better represent the intrinsic spectrum. These results indicate that enhanced persistent emission likely masks the intrinsic characteristics of the burst spectrum, making the data unsuitable for precise mass-radius measurements.

We checked if the relative NS luminosity, $L/L_{\mathrm{Edd}}$, results for the $f_{\rm a}$-corrected NS atmosphere model were similar to the flux values obtained. In this way,  we were able to check whether the model was self-consistent. The model's $L/L_{\mathrm{Edd}}$ is measured at the NS surface \citep{suleimanov_2012}.
The Eddington luminosity of the surface of the NS is defined as:
\begin{equation}
L_{\mathrm{Edd}}=\frac{4 \pi G M c}{\kappa_{\mathrm{e}}}(1+z) , 
\end{equation}
where $\kappa_{\mathrm{e}}$ is the Thomson scattering opacity. In the case of a He-fueled burst, this becomes:
\begin{equation}
L^{\rm He}_{\mathrm{Edd}} = 2L^{\rm H}_{\mathrm{Edd}}=2.52 \times 10^{38} \frac{M}{M_{\odot}} (1 + z) \   \mathrm{erg} \ \mathrm{s}^{-1} . 
\end{equation}
The gravitational redshift is related to the NS parameters as:
\begin{equation}
1+z=\left(1-2 G M / c^2 R\right)^{-1 /2}  . 
\end{equation}
The luminosity at the surface $L$ can then be related to the observed luminosity $L_{\mathrm{obs}}$ as follows:
\begin{equation}
L_{\mathrm{obs}}=L\ (1+z)^{-2},
\end{equation}
and the observed luminosity can be estimated from the observed flux:
\begin{equation}
L_{\mathrm{obs}}=4 \pi D^2 F_{\mathrm{obs}} , 
\end{equation}
where $D$ is the distance.

Assuming a NS of $M=1.4 M_{\odot}$, and using the observed flux from the first second of the burst \#6 and 10.7 kpc for the distance \citep{Baumgard_2021}, we obtain $L/L_{\mathrm{E d d}}$ $\sim$ 0.66. The results, which match well with the values observed in Fig.~\ref{fig:burstatmo_fa}, show the model's estimation makes physical sense. These sub-Eddington luminosities also reflect the lack of observed PRE from any of the bursts.

Finally, it is worth noting that this model assumes the NS does not rotate. However, NSs in LMXBs are typically rapidly rotating  \citep[see][for a review]{galloway_2021}. 
Rapid rotation causes the NS shape to be oblate, leading to variations of the local gravity, the emitted flux, the redshift, and the Doppler shift on latitude. 
All these effects tend to broaden the spectrum \citep{Baubock2015,Suleimanov_2020}. 
Therefore, the possibility that the deviations from the blackbody model are due to the effects of a rapidly rotating NS cannot be ruled out and warrants further investigation. Although addressing this is beyond the scope of this work, it offers a basis for future efforts to develop and test models for rotating NS atmospheres.

\subsection{X-ray burst fuel} 

To study the fuel composition, we calculated the ratio of the persistent fluence between bursts to the total burst fluence: 
\begin{equation}
\alpha= \frac{F_{\mathrm{per}} \Delta T_{\mathrm{rec}} }{E_{\mathrm{b}}}  . 
\end{equation}
We determined that the recurrence timescale, $\Delta T_{\mathrm{rec}}$, is the time between the two consecutive bursts shown in Fig.~\ref{fig:doubleLC}, which is $\sim$ 12.5~h. There are no previous constraints on the recurrence time of \mbox{M15 X-2} in the literature, except for a lower limit of 1.9~d proposed by \citet{white_2001}.

The persistent flux was then obtained from our spectral analysis of the persistent emission before burst \#3 and the total burst fluence was calculated by integrating all the flux values throughout burst \#3. To account for the remaining tail of the flux beyond the end of the pointing, we integrated Eq.~\eqref{eq:lc_model} using its fitted parameters from the end of the exposure to infinity. However, this additional contribution was negligible, falling within the margin of error. 
We ultimately determined $\alpha = 119 \pm 8$. These values indicate a He-rich burst \citep{lewin_993}.
This is consistent with the rapid burst rise and decay times listed in  Table~\ref{tab:lcs}, which are characteristic of helium and soft-state bursts \citep{galloway_2021}. 
The fuel composition is compatible with UCXB systems, defined by their low hydrogen content. UCXBs have three potential donor channels: the white dwarf (WD) channel, the helium star channel, and the evolved main-sequence star channel  \citep[see][for a review]{nelemans_2010}. 

Our findings are in agreement with \citet{dieball_2005}, who found ultraviolet (FUV) evidence of C and/or He lines, but do not constrain the donor any further. Later studies seem to favor a WD companion: \citet{koliopanos_2020} analyzed the X-ray spectrum of \mbox{M15 X-2} and noted the absence of  Fe K$\alpha$ emission, suggesting a C/O or O/Ne/Mg WD donor. \citet{ArmasPadilla_2023} cataloged all known UCXBs, and from the statistical analysis, proposed globular cluster sources usually follow the WD channel and galactic field sources the helium star channel. Typical WDs have an outer layer of residual helium comprising around 1\%  of the star's total mass \citep{Saumon_2022}. In the case of hybrid He-CO WDs, this percentage rises to 2--25\% \citep{Zenati_2019}. Thus, the presence of helium does not have to be fully discarded in the case of a WD companion, but the exact nature of the donor remains unclear. 

The determination of the fuel composition, together with more precise distance measurements, could potentially allow us to reprocess historical bursts from this source with greater accuracy, such as those observed by the \textit{Ginga} satellite \citep{vanParadijs_1990}. 

\section{Conclusions}
\label{section:conclusions}

Using {NICER} observations, we detected seven X-ray bursts from \mbox{M15 X-2} during its 2022 outburst. We obtained light curves of the outburst and the individual bursts and performed X-ray spectroscopy of the persistent emission and time-resolved X-ray spectroscopy of the bursts. Below, we discuss the main conclusions of the study.

\begin{enumerate}
\item All the bursts have very similar morphology, with quick rises of under a second, decays of tens of seconds, and no evidence of PRE. 
The spectra of the persistent emission indicate all of the bursts occurred in the soft state.
\item Time-resolved spectroscopy showed all the bursts exhibit enhanced persistent emission in the first few seconds. This behavior was adequately modeled using the $f_{\rm a}$ factor method.

\item NS atmosphere models did not significantly improve the fits in comparison to the blackbody model. This suggests that enhanced persistent emission likely masks the intrinsic characteristics of the burst spectrum, making the data unsuitable for precise mass-radius measurements.
\item Two consecutive bursts allowed us to derive a recurrence time of 12.5~h. Using this recurrence, we estimated the fuel composition by calculating the alpha parameter and obtained $\alpha = 119 \pm 8$, which is consistent with a He-rich burst. This result is in accordance with the UCXB nature of the source. 
\end{enumerate}

\begin{acknowledgements}
We thank the anonymous referee for their feedback that helped improve the manuscript. 
MADT acknowledges support from the EDUFI Fellowship and the Johannes Andersen Student Programme at the Nordic Optical Telescope. 
We acknowledge support from ESA through the ESA Space Science Faculty Visitor scheme (ESA-SCI-SC-LE-098 and ESA-SCI-SC-LE-203). M.N. is a Fonds de Recherche du Quebec – Nature et Technologies (FRQNT) postdoctoral fellow.
       
\end{acknowledgements}

\bibliographystyle{aa} 
\bibliography{M15new}

\begin{appendix} 
\onecolumn


\section{Additional figures}
\label{app:1}


\begin{figure} [!h]
\centering
\includegraphics[width=0.30\textwidth]{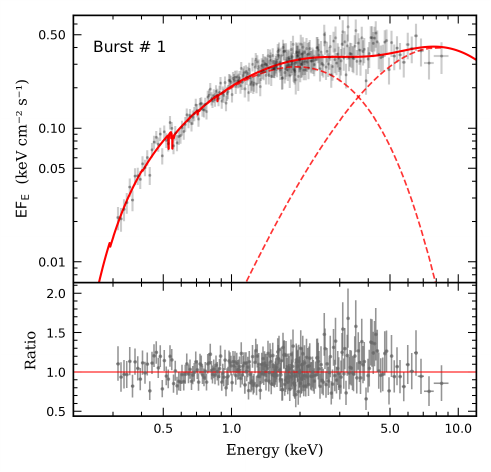}   
\includegraphics[width=0.30\textwidth]{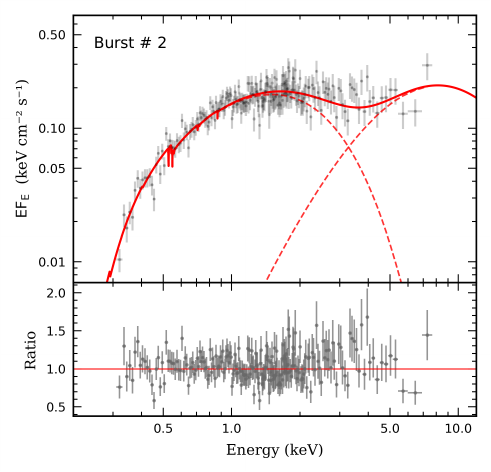}  
\includegraphics[width=0.30\textwidth]{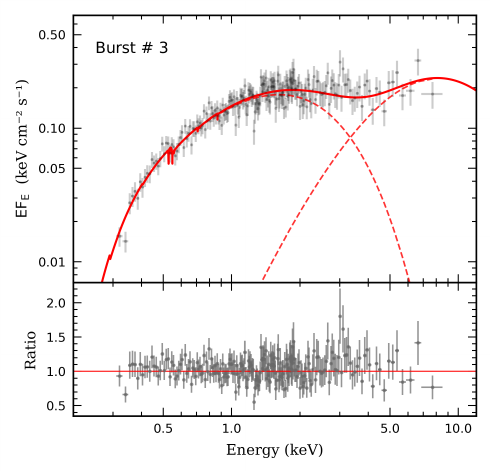}  \\
\includegraphics[width=0.30\textwidth]{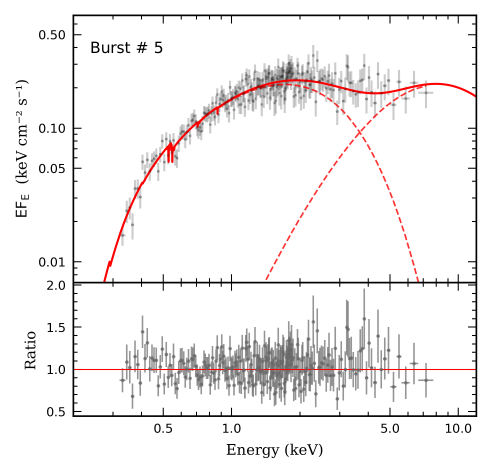}  
\includegraphics[width=0.30\textwidth]{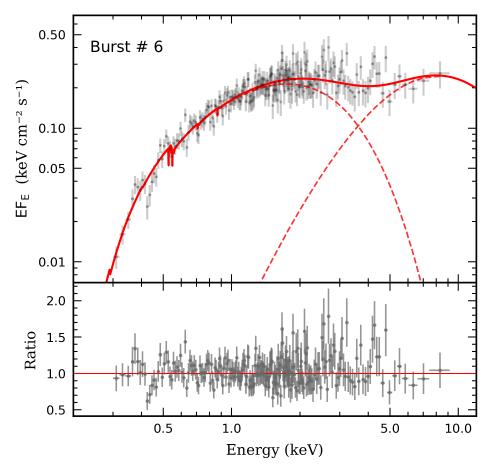}  
\includegraphics[width=0.30\textwidth]{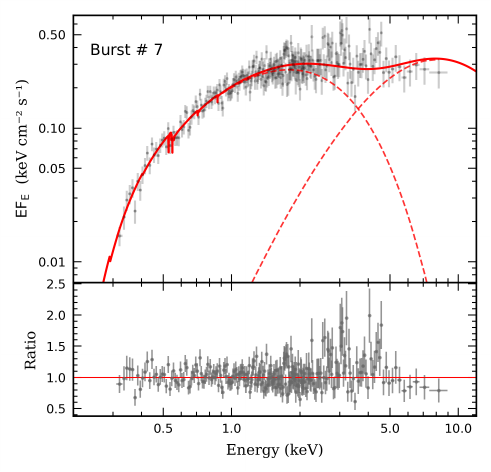}  
\caption{Same as Fig. \ref{fig:PreBurst_spectrum} but for different bursts labeled on the figure. }
\label{fig:PreBurst_spectrum1}
\end{figure}

\begin{landscape}


\begin{figure}  [!h]
\centering
\includegraphics[width=0.24\textwidth]{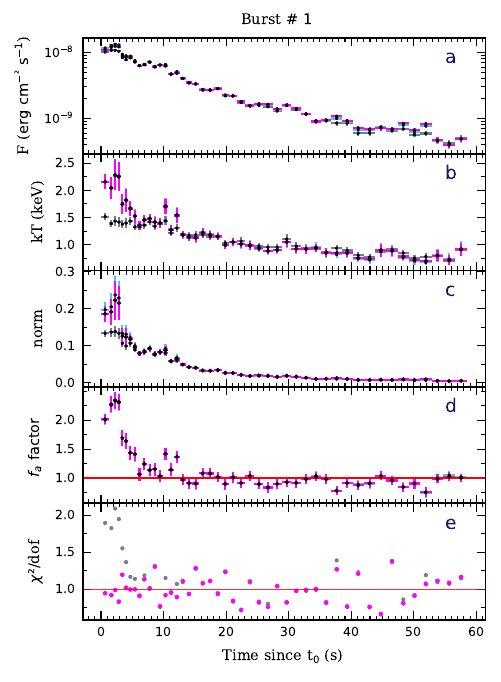}
\includegraphics[width=0.24\textwidth]{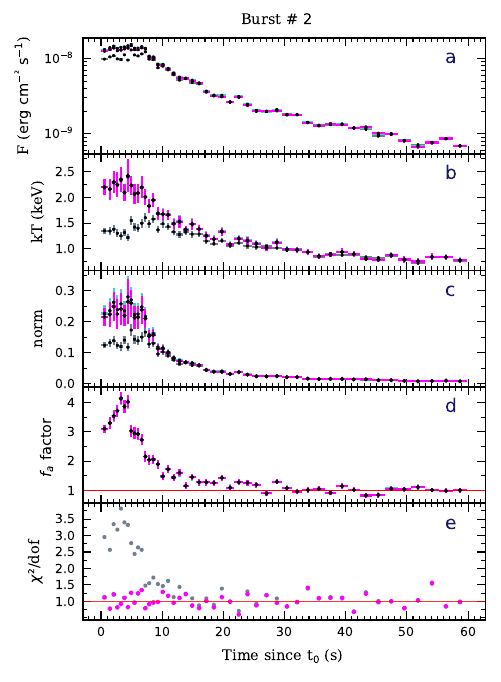}
\includegraphics[width=0.24\textwidth]{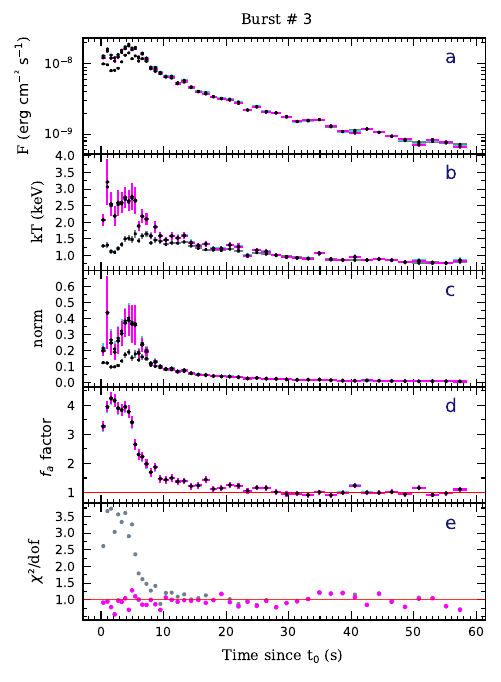}
\includegraphics[width=0.24\textwidth]{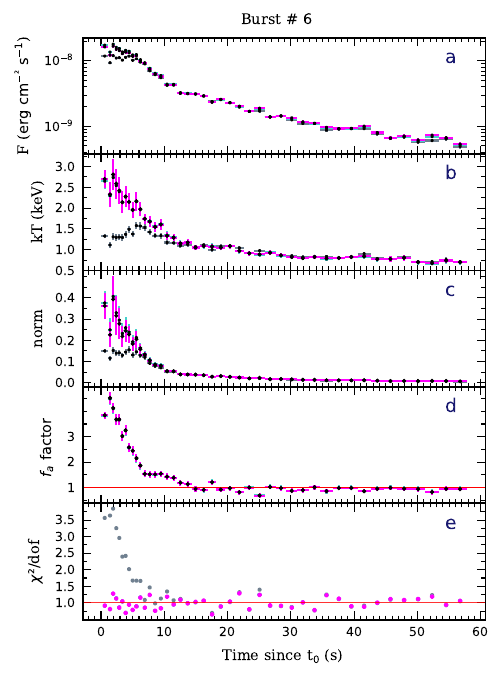}
\includegraphics[width=0.24\textwidth]{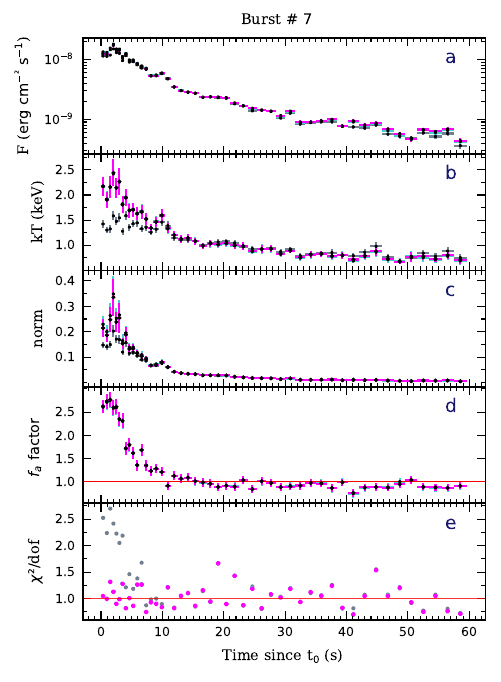}
\caption{Same as Fig. \ref{fig:burstatmo} but for different bursts labeled on the figure.}
\label{fig:blackbody_fa1}
\end{figure} 


\begin{figure} [h!] 
\centering
\includegraphics[width=0.24\textwidth]{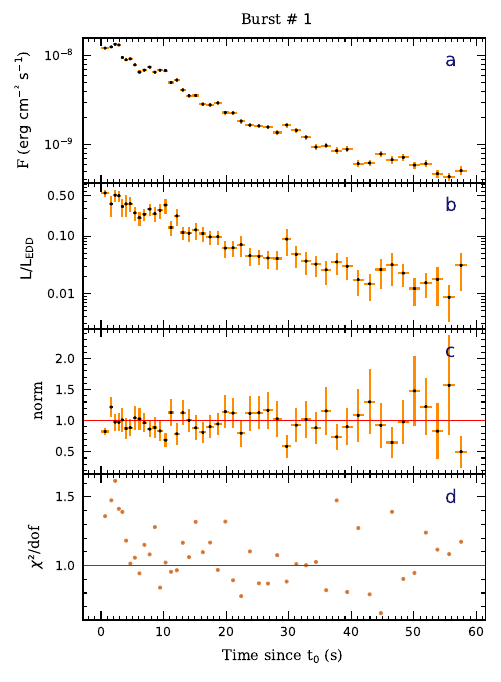}
\includegraphics[width=0.24\textwidth]{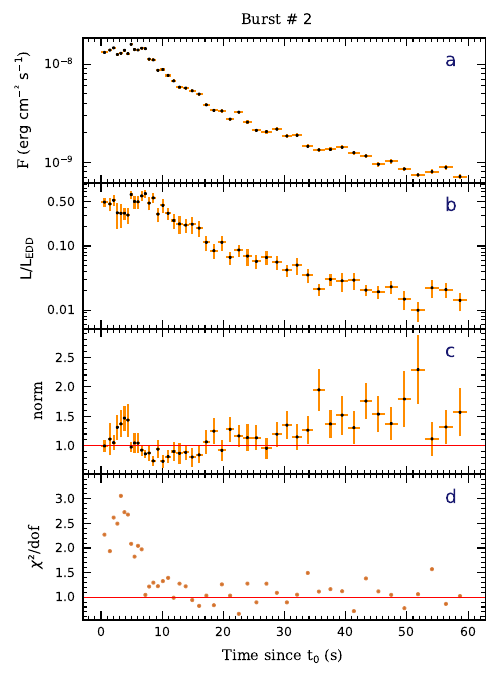}
\includegraphics[width=0.24\textwidth]{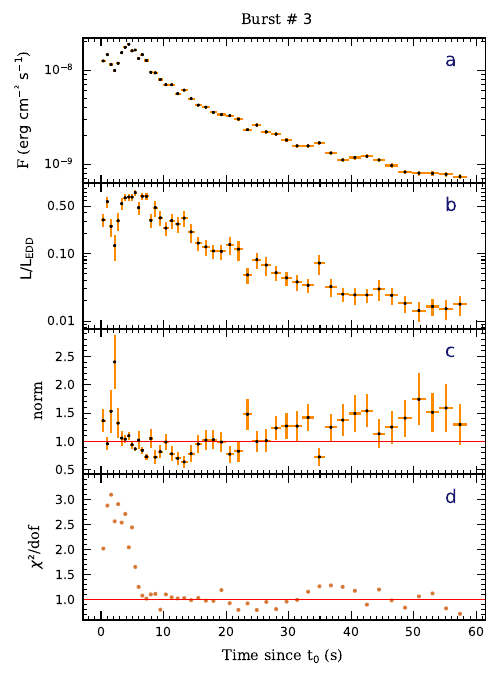}
\includegraphics[width=0.24\textwidth]{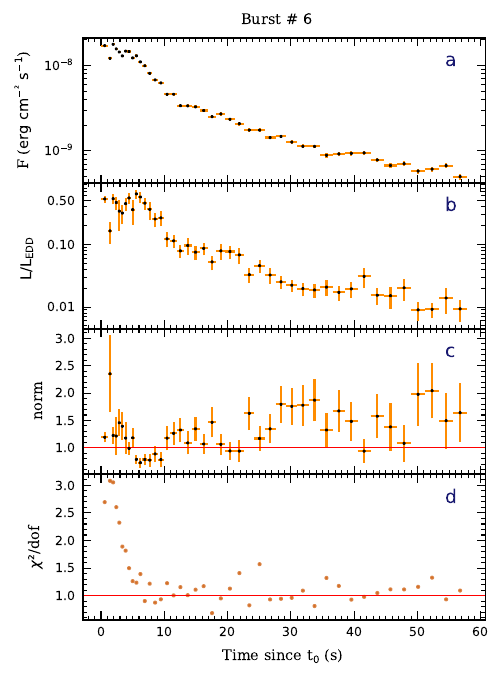}
\includegraphics[width=0.24\textwidth]{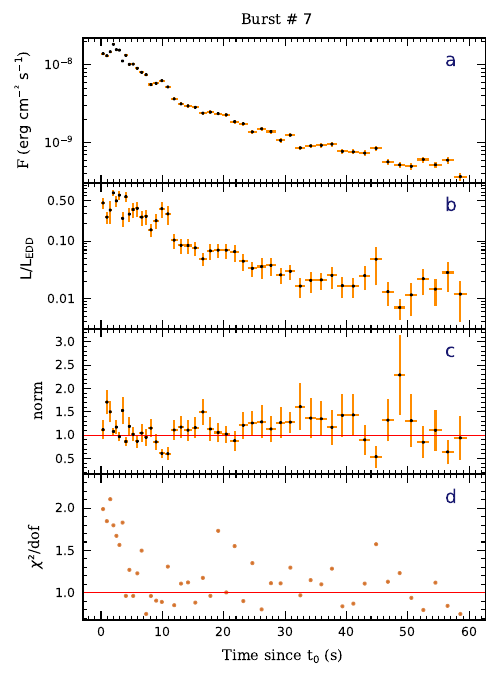}
\caption{Same as Fig. \ref{fig:burstatmo} but for different bursts labeled on the figure.}
\label{fig:burstatmo1}
\end{figure}
  

\begin{figure}  
\centering
\includegraphics[width=0.24\textwidth]{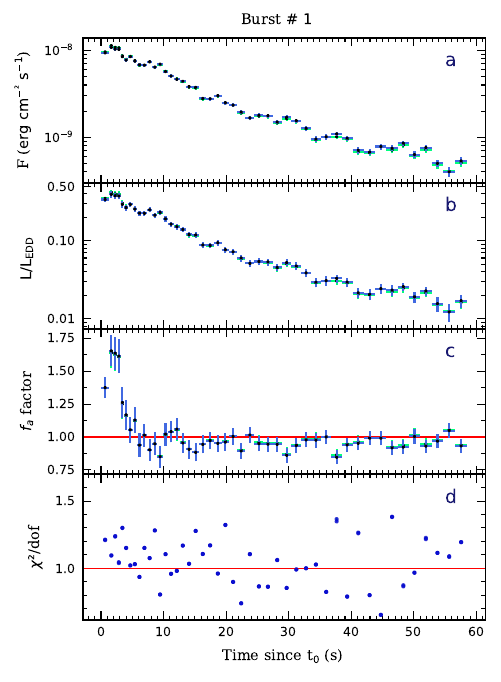}
\includegraphics[width=0.24\textwidth]{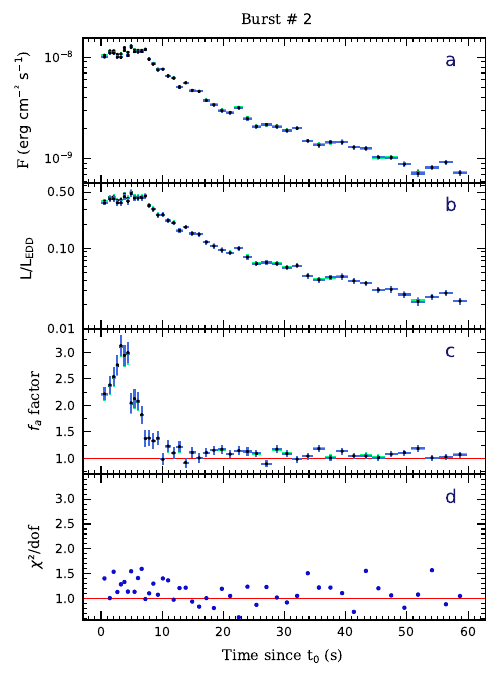}
\includegraphics[width=0.24\textwidth]{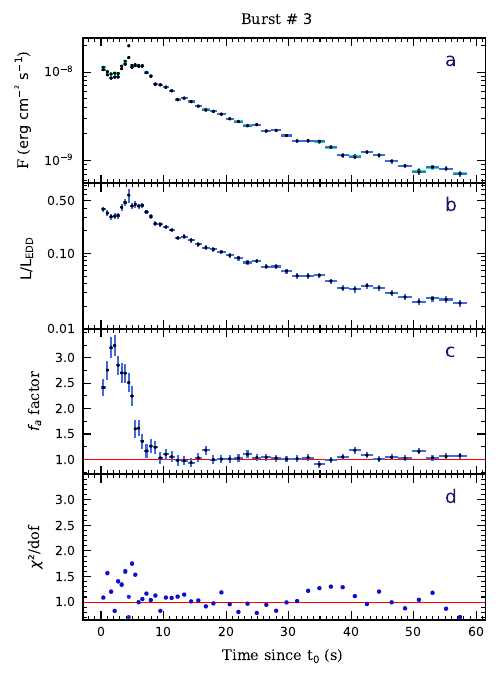}
\includegraphics[width=0.24\textwidth]{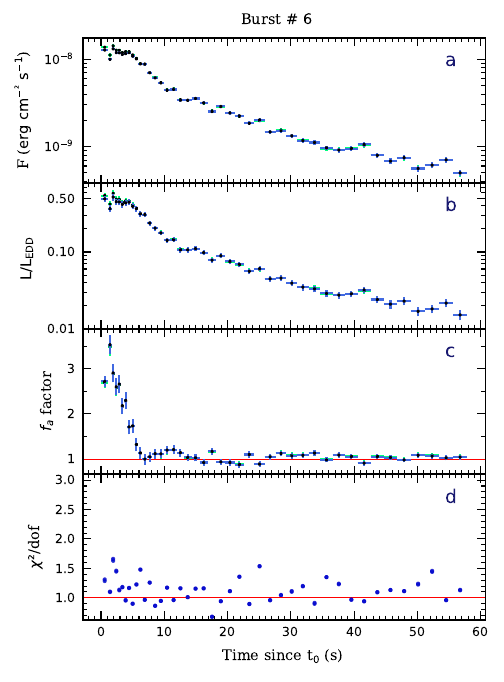}
\includegraphics[width=0.24\textwidth]{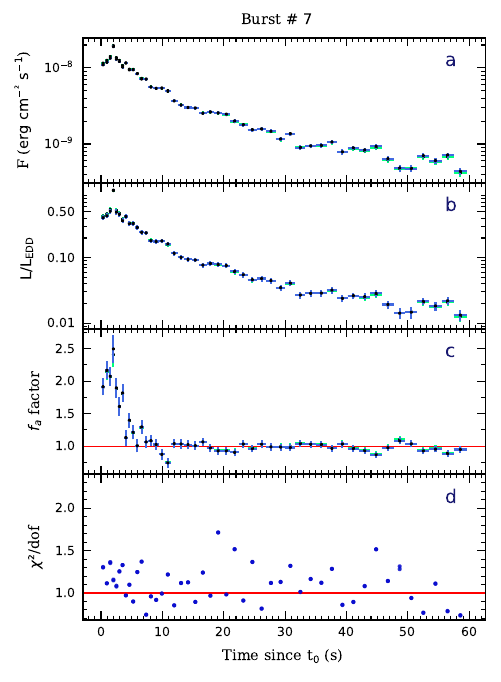}
\caption{Same as Fig. \ref{fig:burstatmo_fa} but for different bursts labeled on the figure.}
\label{fig:burstatmo_fa1}
\end{figure}


\begin{figure} 
\centering
\includegraphics[width=0.24\textwidth]{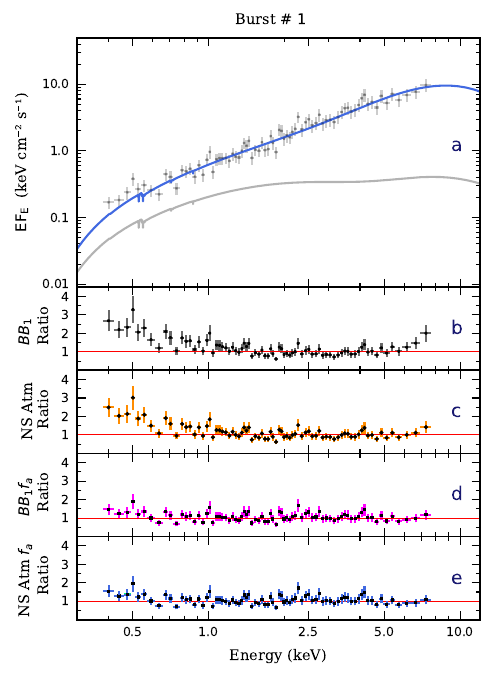}
\includegraphics[width=0.24\textwidth]{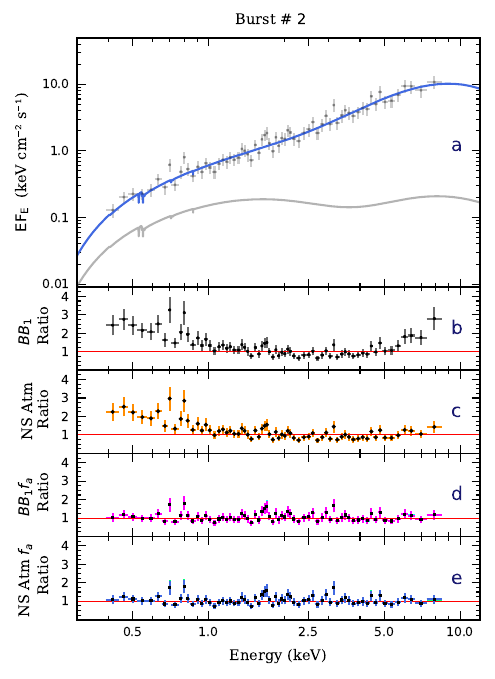}
\includegraphics[width=0.24\textwidth]{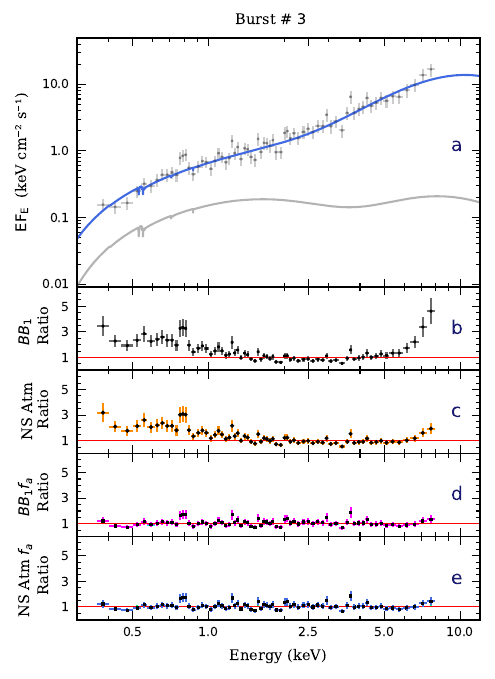}
\includegraphics[width=0.24\textwidth]{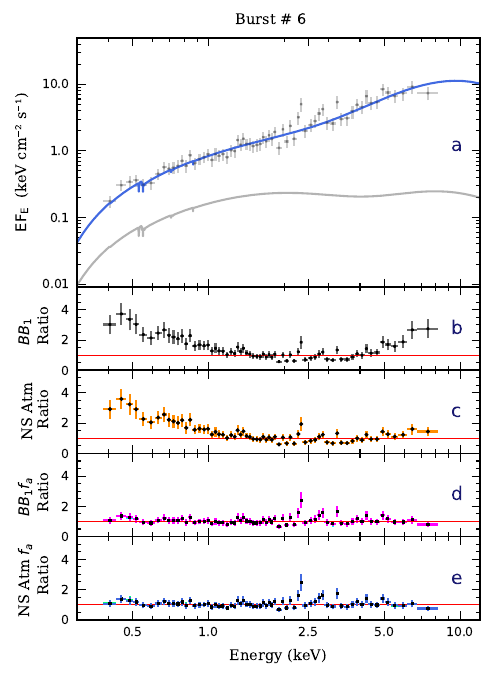}
\includegraphics[width=0.24\textwidth]{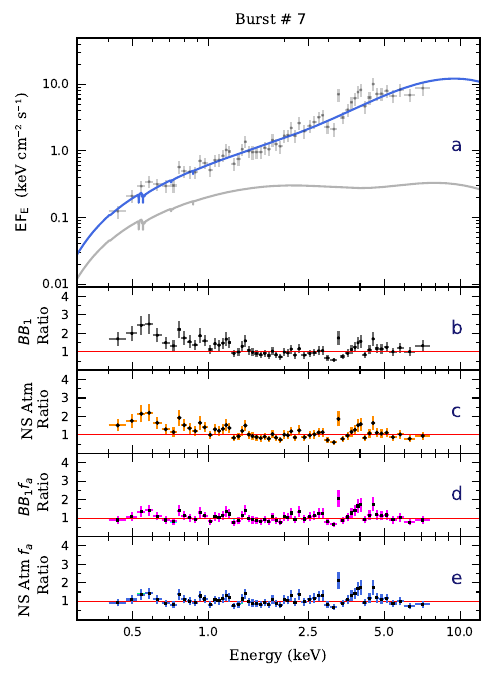}
\caption{Same as Fig. \ref{fig:Residuals} but for different bursts labeled on the figure. }
\label{fig:Residuals1}
\end{figure}
\end{landscape}

\end{appendix}

\end{document}